\documentclass[12pt]{article}
\usepackage{amsmath}
\usepackage{epsfig}
\usepackage{latexsym}
\usepackage{a4wide}
\usepackage{bbm}
\usepackage{feynmf}
\newcommand{\I}{\text{i}}
\newcommand{\E}{\text{e}}

\newcommand{\case}[2]{\frac{\scriptstyle #1}{\scriptstyle #2}}
\newcommand{\re}[1]{(\ref{#1})}
\newcommand{\GBN}{G_{\text{BN}}}
\newcommand{\xiT}{\xi_{\text{T}}}
\renewcommand{\vec}[1]{\boldsymbol{#1}}
\begin{document}
\abovedisplayskip17pt plus2pt minus4pt
\abovedisplayshortskip14pt plus2pt minus4pt
\belowdisplayskip17pt plus2pt minus4pt
\belowdisplayshortskip14pt plus2pt minus4pt
\title{Bloch-Nordsieck Approximation in Linearized Quantum
  Gravity\thanks{Contribution to the proceedings of the ``E.S.~Fradkin
    Memorial Conference'', Moscow June 5-10, 2000}}
\author{Walter Dittrich\\
  Institut f\"ur theoretische Physik, Universit\"at T\"ubingen,\\
  72076 T\"ubingen, Germany}
\maketitle
\begin{abstract}
  The aim of this article is to review Fradkin's contribution in the
  realm of eikonal physics. In particular, the so-called Fradkin
  representation is employed to investigate a certain subclass of
  Feynman diagrams resulting in an expression for the scattering
  amplitude at Planckian energies. The 't Hooft poles are reproduced. 
\end{abstract}

\section{Introduction}
It is not difficult for me to give a tribute to E.S.~Fradkin's
scientific achievements because I already became acquainted with his work
during the second half of the sixties. I was then a graduate
student at Brown University, where Herb Fried lectured in his quantum
field theory course on the Bloch-Nordsieck approximation and the
Schwinger-Fradkin representation. The representation invented by
Fradkin first appeared in a 1966 Nuclear Physics article \cite{1}. It
became the backbone of many of the eikonal articles in the years to
come. The state of the art has been extensively reviewed in
Ref. \cite{2}. Similar methods have been adopted recently in \cite{3}
in order to study Planckian-energy gravitational scattering. I will
pick up the story in 1970, when I published a short article in
Phys. Rev. \cite{4} which is concerned with the summation of a certain
subclass of Feynman diagrams within the Bloch-Nordsieck (or eikonal)
approximation. While this approximation was used in scalar and spinor
QED at that time \cite{5}, it was not until 1992 that the same methods
were applied again to linearized quantum gravity \cite{6}. Now path
integral techniques were en vogue and had long since overtaken
functional differentiation methods as advocated by Schwinger-Symanzik
or Fradkin himself. 

However, in the sequel I will demonstrate within linearized quantum
gravity that there is no need to go beyond the functional
differentiation techniques that were already available in the sixties.

\section{Linearized Quantum Gravity with Spin-$\frac{1}{2}$ Matter
  Interaction}

At the beginning of this chapter we write down the linearized
Einstein-Hilbert action coupled to a spin-$\frac{1}{2}$ matter
field. After the metric $g_{\mu\nu}$ has been expanded about a
Minkowski flat background, $g_{\mu\nu}=\eta_{\mu\nu}+h_{\mu\nu}$, the
gauge-fixed linearized action is given by
\begin{equation}
S=\frac{1}{16\pi G} \int d^4x \, \frac{1}{2} \left[ \frac{1}{2}
  h_{\mu\nu} \partial^2 h^{\mu\nu} -\frac{1}{4} h \partial^2
  h+16\pi G\,T^{\mu\nu} h_{\mu\nu} \right]. \label{1}
\end{equation}
In the gauge-fixing term $-\frac{1}{2} C_\mu^2$ we have chosen the
DeDonder gauge $C_\mu=\partial_\nu h^\nu_\mu -\frac{1}{2} \partial_\mu
h$, where $h=h^\lambda_\lambda$ 
and all indices are to be raised and lowered with
$\eta_{\mu\nu}=\text{diag}(-1,+1,+1,+1)$. As usual we set
$\hbar=1=c$. The linearized matter action in Eq. \re{1} makes use of
\begin{equation}
T_{\mu\nu}=\frac{1}{4} \left\{ \bar{\psi} \gamma_\mu \Bigl(
    \case{1}{\I} \partial_\nu \psi\Bigr) -\Bigl(
    \case{1}{\I} \partial_\nu \bar{\psi}\Bigr)\gamma_\mu \psi +\bigl(
    \mu\leftrightarrow \nu\bigr) \right\}. \label{2}
\end{equation}
Note that the gravitational part in Eq. \re{1} leads to the graviton
propagator
\begin{equation}
D_{\mu\nu\lambda\sigma}(k) =\frac{16\pi G}{k^2 -\I \epsilon} \bigl(
\eta_{\mu\lambda} \eta_{\nu\sigma} +\eta_{\mu\sigma} \eta_{\nu\lambda}
-\eta_{\mu\nu} \eta_{\lambda\sigma} \bigr). \label{3}
\end{equation}
Writing the Lagrangian of the matter interaction as
\begin{displaymath}
{\cal L}(\psi,\bar{\psi}, h_{\mu\nu})=-\bar{\psi} \Bigl( \gamma
  \case{1}{\I} \partial +m \Bigr)\psi +\frac{1}{2} h^{\mu\nu}
  T_{\mu\nu}, 
\end{displaymath}
we obtain the Green's function equation from the action principle,
\begin{equation}
\left[ m+\gamma \frac{1}{\I} \partial^x -\frac{1}{4} \gamma_\mu \left(
      \frac{1}{\I} \partial^x_\nu h^{\mu\nu}(x) +h^{\mu\nu}(x)
      \frac{1}{\I} \partial^x_\nu \right) \right] G(x,y|h_{\mu\nu}
      )=\delta(x-y). \label{4}
\end{equation}
There are several methods to solve Eq. \re{4} approximately. Since we
are interested in elastic forward scattering processes with the
exchange of an arbitrary number of soft gravitons, we proceed to
evaluate Eq. \re{4} using the Bloch-Nordsieck (BN) or eikonal
approximation. For this reason we begin by performing a semi-Fourier
transform,
\begin{equation}
G(x,y|h_{\mu\nu}) =\int \frac{d^4p}{(2\pi)^4}\, \E^{\I px}\,
G(p,y|h). \label{5}
\end{equation}
In this way we obtain the following expression:
\begin{equation}
\left[ m+\gamma\frac{1}{\I} \partial^x \right]G(x,y|h) =\int
\frac{d^4p}{(2\pi)^4}\, \E^{\I px} \left( \E^{-\I py} + \frac{1}{2}
  \gamma_\mu h^{\mu\nu}(x) p_\nu \, G(p,y|h)\right). \label{6}
\end{equation}
In this equation we have dropped the term $\frac{1}{4} \gamma_\mu
\frac{1}{\I} \partial^x_\nu h^{\mu\nu}(x)=\frac{1}{8} \gamma_\mu
\frac{1}{\I} \partial^\mu h(x)$ which leads to a kind of self-coupling
-- the same index $\mu$.

When we now perform the BN-approximation in Minkowski flat space by
setting 
\begin{eqnarray} 
\gamma_\mu &\to& \langle \gamma_\mu\rangle =v_\mu
  =\frac{p_\mu}{m}, \qquad v^2=-1, \label{7} \\
p_\nu&\simeq& p_\nu'\simeq \frac{p_\nu+p_\nu'}{2}, \nonumber
\end{eqnarray}
we observe that it is exactly the second term on the right-hand side
of Eq. \re{6} that gives us the desired $p_\mu h^{\mu\nu}
p_\nu$-interaction term. Within this approximate treatment we now have
to solve the Green's function equation, 
\begin{equation}
\left[ m+v \frac{1}{\I} \partial -\frac{1}{2} v_\mu p_\nu h^{\mu\nu}
  (x) \right] \GBN(x,y|h) =\delta(x-y). \label{8}
\end{equation}
At this stage we first introduce Schwinger's proper-time
representation \cite{7} (we follow the pattern of calculations as
outlined in Ref. \cite{8} for a scalar field theory):
\begin{eqnarray}
\GBN[h] &=& \I \int\limits_0^\infty ds \, \E^{-\I s\left[m-\I\epsilon
    +v\frac{1}{\I} \partial +B\right]},
    \qquad B:=-\frac{1}{2} v_\mu p_\nu h^{\mu\nu} \nonumber\\
&=&\I \int\limits_0^\infty ds \, \E^{-\I s(m-\I\epsilon)} \E^{\I s[v\I
    \partial -B]}=\I \int\limits_0^\infty ds\, \E^{-\I s(m-\I
    \epsilon)} U(s) , \label{9}
\end{eqnarray}
where $U(s)=\E^{\I s[v\I\partial -B]}$. 
In the second step we follow Fradkin by coupling the operator
$(v\cdot\I \partial)$ to the ``source'' $\nu(s)$. This is done by
introducing
\begin{equation} 
U(s,\nu)=T_{s'} \left(
  \E^{\I\int\limits_0^s ds'\, \left( v\cdot\I\partial- B+\nu(s')
  (v\cdot\I\partial)\right)}\right) \label{10}
\end{equation}
with the property $U(s,\nu)|_{\nu\to 0} =U(s)$. 
If we now recall the simple formula
\begin{displaymath}
F\left\{ \frac{\delta}{\delta\nu}\right\} \, \E^{\I\int\nu f}
=F\{ \I f\} \, \E^{\I\int \nu f} 
\end{displaymath}
or
\begin{equation}
\E^{\int\limits_0^s ds'\, \frac{\delta}{\delta \nu(s')}} \, \E^{\I
  \int\limits_0^s ds' \, \nu(s') (v\cdot\I\partial)} 
= \E^{\I \int\limits_0^s ds'\, (v\cdot\I\partial)} \,
  \E^{\I\int\limits_0^s ds' \nu(s') (v\cdot\I\partial)}, \label{11}
\end{equation}
we obtain
\begin{equation}
U(s,\nu) =\mathbbm{1}\, \E^{\int\limits_0^{s-\epsilon} ds'\,
  \frac{\delta}{\delta \nu(s')}}\, T_{s'} \left( \E^{\int\limits_0^s
  ds' \bigl[ -\nu(s') (v\cdot \partial) -\I B\bigr]}
  \right). \label{12}
\end{equation}
The expressions \re{10} and \re{12} imply
\begin{eqnarray}
\frac{\partial U(s,\nu)}{\partial s} &=& \I \Bigl[ v\cdot \I\partial -B
+\nu(s) (v\cdot\I\partial) \Bigr] U(s,\nu), \qquad U(0,0)=1\label{13}\\
\text{and}\qquad \frac{1}{\I} \frac{\delta U(s,\nu)}{\delta \nu(s)} &=&
(v\cdot\I\partial) U(s,\nu). \label{14}
\end{eqnarray}
At this point it is convenient to define two new functions $W(s)$ and
$f(s)$ according to
\begin{eqnarray}
W(s)&\equiv& T_{s'}\left( \E^{\int\limits_0^s ds'\,
    [-\nu(s')(v\cdot\partial) -\I B]} \right), \nonumber\\
W(s) &=&\E^{- \int\limits_0^s ds'\, \nu(s') (v\cdot \partial)} \,
f(s), \label{15} \\
\text{with} \qquad f(s) &=&\E^{ \int\limits_0^s ds'\, \nu(s') (v\cdot
  \partial)} \, W(s). \label{16} 
\end{eqnarray}
It is then easy to set up a differential equation for $\frac{\partial
  f}{\partial s}$ with $f(0)=1$ whose solution is given by
\begin{equation}
\langle x| f(s)| y\rangle =\E^{-\I\int\limits_0^s ds'\, B\left(
    x+\int\limits_0^{s'} ds''\, \nu(s'')\, v\right)}\, \langle
    x|y\rangle. \label{17}
\end{equation}
When this expression is substituted into Eq. \re{15} we obtain 
\begin{equation}
\langle x|W(s)|y\rangle=\E^{-\I\int\limits_0^s ds'\,
  B\left(y+\int\limits_0^{s'} ds''\, \nu(s'') v\right)}
  \delta\left(x-y-\int\limits_0^s ds'\, \nu(s') v\right). \label{18}
\end{equation}
This result enables us to write our solution in the form
\begin{eqnarray}
\GBN(x,y|h) &=&\I\int\limits_0^\infty ds\, \E^{-\I sm}
\E^{\int\limits_0^s ds'\, \frac{\delta}{\delta \nu(s')}} \E^{\I
  \frac{p_\mu p_\nu}{2m} \int\limits_0^s ds'\, h_{\mu\nu} \left(
    y+\int\limits_0^{s'} ds''\, \nu(s'') v\right)} \nonumber\\
&&\qquad\qquad\times\delta\left(
  x-y-\int\limits_0^s ds'\, \nu(s')
  v\right)\Biggr|_{\nu=0}. \label{19}
\end{eqnarray}
A simple functional Fourier transform yields
\begin{displaymath}
\E^{\int\limits_0^sds'\, \frac{\delta}{\delta\nu(s')}} F\bigl\{ \nu
\bigr\}\Bigr|_{\nu=0} ={\cal N} \int {\cal D}\phi\,
F\bigl\{\phi\bigr\} \, \E^{\I\int\limits_0^s ds'\, \phi(s')}, 
\end{displaymath}
so that an equivalent path integral representation of Eq. \re{19} is
given by
\begin{eqnarray}
\GBN(x,y|h) &=&\I\int\limits_0^\infty ds \,\E^{-\I sm}
{\cal N}(s) \int {\cal D}\phi\, \E^{\I \int\limits_0^s ds'
  \, \phi(s')}  \E^{\I \frac{p_\mu p_\nu}{2m} \int\limits_0^s ds'\,
  h_{\mu\nu} \left( y+\int\limits_0^{s'} ds''\, \phi(s'') v\right)}
\nonumber\\
&&\qquad\qquad\times  \delta\left(  x-y-\int\limits_0^s ds'\, \phi(s')
  v\right), \label{20}
\end{eqnarray}
with ${\cal N}(s)^{-1} =\int{\cal D} \phi\, \exp\left(\I \int_0^s ds'
  \, \phi(s')\right)$. For similar formulas in QED, see Ref. \cite{9}.

If we now, in the eikonal spirit, restrict the sum over paths to
straight-line paths between $y$ and $x$, we obtain from Eq. \re{19} 
\begin{equation}
\GBN(x,y|h)=\int \frac{d^4p}{(2\pi)^4}\, \E^{\I p(x-y)} \I
\int\limits_0^\infty ds\, \E^{-\I s(m+vp)} \E^{\I \frac{p^\mu
    p^\nu}{2m} \int\limits_0^sds'\, h_{\mu\nu} \left[ x \frac{s'}{s}
    +\left( 1-\frac{s'}{s} \right) y\right]}. \label{21}
\end{equation}
Introducing the variable $u=\frac{s'}{s}$, the $s$ integration in
Eq. \re{21} is easily performed:
\begin{equation} 
\int \frac{d^4p}{(2\pi)^4}\, \E^{\I p(x-y)}
\left\{\left[ m-\frac{p^\mu
    p^\nu}{2m} \int\limits_0^1 du\, h_{\mu\nu} (xu+(1-u)y)\right]
+vp\right\}^{-1}. \label{22}
\end{equation}
This representation, compact as it is, is, however, difficult to use
in momentum space. Fortunately Eq. \re{21} can also be rewritten in
the form
\begin{equation}
\GBN(x,y|h)=\int \frac{d^4p}{(2\pi)^4}\, \E^{\I p(x-y)} \, \I \!
\int\limits_0^\infty ds\, \E^{-\I s(m+vp)}
\E^{\frac{\I}{2} \frac{p^\mu  p^\nu}{m} \int\limits_0^s ds'\,
  h_{\mu\nu} \left( x -\frac{p}{m}s' \right) }. \label{23}
\end{equation}
It is this representation of the eikonalized Green's function which we
will employ in the sequel when computing scattering amplitudes.

\section{Green's Function, Newton Potential}

In this section we are interested in the four-point Green's function
where we treat the two-point Green's function for particle \#1 in the
BN-approximation while the Green's function for particle \#2 is the
exact one, i.e., satisfies Eq. \re{4}. At this point we need the
on-shell truncated, connected Green's function for particle \#1:
\begin{equation}
\GBN^{\text{c}}(p,p'|h)=\lim_{m+vp=0,\, m+vp'=0} (m+vp)\,
\GBN(p,p'|h)\,(m+vp'), \label{24}
\end{equation}
where the momentum-space version of Eq. \re{23} is given by
\begin{equation}
\GBN(p,p'|h)=\int d^4x\, \E^{-\I (p-p')x} \,\I\!
\int\limits_0^\infty ds\, \E^{-\I s(m+vp')}
\E^{\frac{\I}{2} v^\mu  p'{}^\nu \int\limits_{0}^s ds'\,
  h_{\mu\nu} \left( x -v s' \right) }. \label{25}
\end{equation}
Two integrations by parts on the variable $s$ then lead to
\begin{equation}
\GBN^{\text{c}}(p,p'|h)=-\I \int d^4x\, \E^{-\I (p-p')x} 
\frac{\partial}{\partial s} \left( \E^{\I \frac{p^\mu  p^\nu}{2m}
    \int\limits_{-\infty}^s ds'\,  h_{\mu\nu} 
   \left( x +\frac{p}{m}s' \right)} \right)_{s=0}. \label{26}
\end{equation}
Making a change of variables,
\begin{displaymath}
x=z+\frac{p}{m} \beta,
\end{displaymath}
where $z$ is space-like ($-z^2<0$) and $p$ time-like ($-p^2>0$), we
find
\begin{equation}
\GBN^{\text{c}}(p,p'|h)=-\I \frac{p^0}{m} \int d^3\mathbf{z}\,
\E^{\I(p'-p) z}\, \E^{\I \frac{p^\mu p^\nu}{2m}
  \int\limits_{-\infty}^\infty d\tau\, h_{\mu\nu}\left( z+\frac{p}{m}
    \tau \right)}. \label{27}
\end{equation}
Since we are dealing with spinors we must not forget to sandwich
Eq. \re{27} between $\bar{u}_\alpha(p) \dots u_\beta(p')\to
\delta_{\alpha\beta}$, so that Eq. \re{27} turns into
\begin{eqnarray}
\GBN^{\text{c}}(p,p'|h)&=&-\I \frac{p^0}{m}\,\delta_{\alpha\beta} \int
d^3\mathbf{z}\, \E^{\I(p'-p) z}\, \E^{\I \frac{p^\mu p^\nu}{2m}
  \int\limits_{-\infty}^\infty d\tau\, h_{\mu\nu}\left( z+\frac{p}{m}
    \tau \right)}, \label{28}
\end{eqnarray}
For further studies it is convenient to rewrite Eq. \re{28} in the
form
\begin{eqnarray}
\langle p,\alpha|\GBN^{\text{c}}[h]| p',\beta\rangle&=&-\I
\frac{p^0}{m}\,\delta_{\alpha\beta} \int 
d^3\mathbf{z}\, \E^{\I(p'-p) z}\, \E^{\frac{\I }{2} \int d^4\xi\,
  T^{\mu\nu}(\xi) h_{\mu\nu}(\xi)}, \label{29}\\
T^{\mu\nu}(\xi)&:=& \int\limits_{-\infty}^\infty d\tau\, \frac{1}{m}\,
p^\mu p^\nu \, \delta^4\left( \xi -\left(z+ \frac{p}{m} \tau
  \right)\right), \label{30}
\end{eqnarray}
where $T^{\mu\nu}$ is the classical energy--momentum tensor for a
point particle with momentum $p$. 

Now let us return to our forward scattering process viewed in a system
where particle \#1 is highly relativistic compared to particle
\#2. Then the truncated scattering amplitude is given by
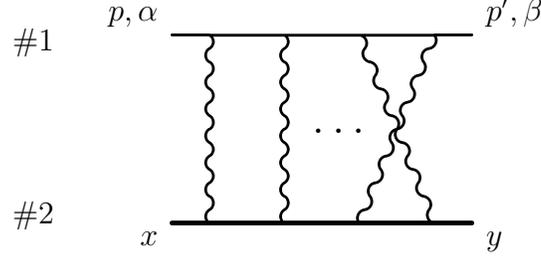
\begin{figure}[h]
\begin{center}
\begin{picture}(100,25)
\put(25,0){
\begin{fmffile}{fmfpicgamma1}
\begin{fmfgraph*}(50,25)
\fmfleft{i1,i2}
\fmflabel{$p,\alpha$}{i2}
\fmflabel{$x$}{i1}
\fmfright{o1,o2}
\fmflabel{$p',\beta$}{o2}
\fmflabel{$y$}{o1}
\fmf{plain,width=15}{i1,v1}
\fmf{plain,tension=.5,width=15}{v1,v2}
\fmf{plain,tension=.5,width=15}{v2,v3}
\fmf{plain,tension=.5,width=15}{v3,v4}
\fmf{plain,width=15}{v4,o1}
\fmf{plain}{i2,v5}
\fmf{plain,tension=.5}{v5,v6}
\fmf{plain,tension=.5}{v6,v7}
\fmf{plain,tension=.5}{v7,v8}
\fmf{plain}{v8,o2}
\fmf{photon,tension=0}{v1,v5}
\fmf{photon,tension=0}{v2,v6}
\fmf{photon,tension=0}{v3,v8}
\fmf{photon,tension=0}{v4,v7}
\end{fmfgraph*}
\end{fmffile}}
\put(13,23){\#1}
\put(13,0){\#2}
\put(51,12){\Large \dots}
\end{picture}
\end{center}
\end{figure}
\begin{eqnarray}
&&\!\!\!\!\!\!\!\!\!\!\!\!
\langle x,p\alpha|y, p'\beta\rangle \equiv G(x,p\alpha;y,p'\beta|h)
\nonumber\\
&&=\E^{-\I \frac{\delta}{\delta h_1^{\alpha\beta}}
    D^{\alpha\beta\gamma\delta} \frac{\delta}{\delta
      h_2^{\gamma\delta}}} \,
  \GBN^{\text{c}}{}_1(p,\alpha,p',\beta|h_1)\,
  G_2(x,y|h_2)\biggl|_{h_{1\mu\nu}=0=h_{2\mu\nu}} \nonumber\\
&&= \E^{-\I \frac{\delta}{\delta h_1^{\alpha\beta}}
    D^{\alpha\beta\gamma\delta} \frac{\delta}{\delta
      h_1^{\gamma\delta}}} \, \int d^4x_1\, \E^{\I(p'-p)x_1}\,
  (-\I\delta_{\alpha\beta}) \nonumber\\
&&\quad\times\frac{\partial}{\partial \tau_1} \left(
    \E^{\I \frac{p^\mu  p^\nu}{2m} \int\limits_{-\infty}^{\tau_1}
      d\tau_1'\, h_{1\mu\nu}  \left( x_1 +\frac{p}{m_1}\tau_1'
      \right)} \right)_{\tau_1=0}
  G_2(x,y|h_2)\Biggl|_{h_{1\mu\nu}=0=h_{2\mu\nu}} . \label{31}
\end{eqnarray}
The final result of the functional differentiation in Eq. \re{31} is
\begin{eqnarray}
&&\!\!\!\!\!\!\!\!\langle p,\alpha|\bigl( \psi(x) \bar{\psi}(y)
\bigr)_+|p', \beta\rangle^{h_{\mu\nu}} =\GBN(x,p\alpha;y,p'\beta|h)
\nonumber\\ 
&&= -\I \frac{p^0}{m}\,\delta_{\alpha\beta} \int d^3\mathbf{z}\, 
\E^{\I(p'-p) z}\, G_2\biggl(x,y\bigg|\underbrace{\frac{p^\mu p^\nu}{2m_1}
    \int\limits_{-\infty}^\infty d\tau\, D^{\mu\nu\alpha\beta} \left(
      z+ \frac{p}{m_1} \tau -\xi \right)}_{=h_1^{\alpha\beta}\left(
      z+\frac{p}{m_1}\tau-\xi\right)} \biggr), \label{32}\\
&&\text{with}\quad h_{1\mu\nu}(\xi) =\frac{1}{2} \int d^4\xi' \,
D_{\mu\nu\alpha\beta}(\xi-\xi') \, T_1^{\alpha\beta}(\xi'). \nonumber
\end{eqnarray}
Formula \re{32} expresses the four-point function in terms of the
Green's function of the spinor particle \#2 in presence of the
linearized gravitational background field $h_{1\mu\nu}$ generated by
the (eikonalized) incoming particle \#1. 

We can rewrite our result \re{32} as
\begin{displaymath}
\GBN=-\I \frac{p^0}{m}\,\delta_{\alpha\beta} \int d^3\mathbf{z}\, 
\E^{\I(p'-p) z}\, G_2\left(x+z,y+z\bigg|\frac{p^\mu p^\nu}{2m_1}
    \int\limits_{-\infty}^\infty d\tau\, D^{\mu\nu\alpha\beta} \left(
      \xi- \frac{p}{m_1} \tau \right)\right).
\end{displaymath}
Using
\begin{displaymath}
D_{\mu\nu\alpha\beta}\left(\xi -\frac{p}{m_1} \tau\right) =16\pi G
\int\frac{d^4 k}{(2\pi)^4} \, \frac{\E^{\I k\left( \xi -\frac{p}{m_1}
      \tau \right)}}{k^2 -\I \epsilon} \, \bigl( \eta_{\mu\alpha}
\eta_{\nu\beta} +\eta_{\mu\beta} \eta_{\nu\alpha} -\eta_{\mu\nu}
\eta_{\alpha\beta} \bigr), 
\end{displaymath}
we find for the functional argument of $G_2$:
\begin{displaymath}
\frac{8\pi G}{m_1} \bigl( 2 p_\alpha p_\beta
-\underbrace{p^2}_{-m_1^2} \eta_{\alpha\beta} \bigr)
\int\limits_{-\infty}^\infty d\tau\, D \left( \xi -\frac{p}{m_1}
  \tau \right),
\end{displaymath}
or: 
\begin{eqnarray}
\GBN(x,p\alpha;y,p'\beta|h)&=& -\I \frac{p^0}{m}\,\delta_{\alpha\beta}
\int d^3\mathbf{z}\,  \E^{\I(p'-p) z}\, \label{33}\\
&&\quad\times G_2\left(x+z,y+z\bigg|
8\pi G \bigl( 2 p_\alpha p_\beta-p^2 \eta_{\alpha\beta} \bigr) 
\int\limits_{-\infty}^\infty d\tau\, D \left( \xi -{p}\tau \right)
\right). \nonumber
\end{eqnarray}
Now let us go to the rest frame of particle \#1: 
\begin{displaymath}
p=(p^0,\mathbf{p}) \equiv (m_1 ,\mathbf{0}). 
\end{displaymath}
Then the functional argument of $G_2$ in Eq. \re{33} turns into
($\eta^{00}=-1$) 
\begin{eqnarray}
&&\!\!\!\!\!\!\!\!\!\!
8\pi G (2m_1^2 -m_1^2)\underbrace{\int\limits_{-\infty}^\infty d\tau\,
D(m_1\tau-\xi^0,\boldsymbol{0}-
\boldsymbol{\xi})}_{\stackrel{m_1\tau\,=\,t}{=}\frac{1}{m_1}
\underbrace{\int\limits_{-\infty}^\infty dt\,
  D(t-\xi^0,-\boldsymbol{\xi})}_{=\frac{1}{4\pi|\boldsymbol{\xi}|},
  \quad |\boldsymbol{\xi}|=r}}, \nonumber\\
&&=\frac{2Gm_1}{r}. \label{34}
\end{eqnarray}
Recall that the functional argument of $G_2$ in Eq. \re{33} is also
the linearized external soft gravitational field $h_1^{\alpha\beta}$:
\begin{displaymath}
h_1^{\alpha\beta} =8\pi G (2p^\alpha p^\beta +m_1^2
\eta^{\alpha\beta}) \int\limits_{-\infty}^\infty d\tau\,
D(\xi-p\tau). 
\end{displaymath}
So we find for 
\begin{equation}
h^{00} =\bigl(h_{00}=\bigr) \, \frac{2Gm}{r}. \label{35}
\end{equation}
Likewise, $h_{jj}=8\pi G m_1^2 \frac{1}{m_1} \left(\frac{1}{4\pi r}
\right)$ or
\begin{equation}
h_{jj} =\frac{2G m_1}{r}, \label{36}
\end{equation}
i.e., if we choose to work in the rest frame of particle \#1, this
gives rise -- in linearized gravity -- to the metric
$g_{\mu\nu}=\eta_{\mu\nu} + h_{\mu\nu}$:
\begin{eqnarray}
(ds)^2&=&g_{\mu\nu}\, dx^\mu\, dx^\nu \nonumber\\
&\stackrel{\text{lin.}}{=}&(-1+h_{00})\, dt^2 +(1+ h_{jj})\,
d\mathbf{x}^2 +\dots, \nonumber
\end{eqnarray}
where we have set $c=1$. So we obtain
\begin{equation}
(ds)^2=\left(-1+\frac{2Gm_1}{r}\right) dt^2 + \left( 1+
  \frac{2Gm_1}{r} \right) \bigl( dr^2+r^2\,d\theta^2+r^2\sin^2
  \theta\, d\phi^2 \bigr) \label{37}
\end{equation} 
instead of the square of the line element in the Schwarzschild metric:
\begin{displaymath}
(ds)^2=-\left( 1-\frac{2Gm_1}{r} \right) dt^2 +
\frac{1}{1-\frac{2Gm_1}{r}}\, dr^2 + r^2 d\Omega^2, 
\end{displaymath}
where we have abbreviated $d\theta^2+\sin^2 \theta\, d\phi^2 $ by
$d\Omega^2$. 

\section{Scattering Amplitude, 't Hooft Poles}

Finally we turn to the elastic scattering of two electrons which
exchange gravitons in all possible ladder-type ways. As long as
self-energy structure and vertex corrections are neglected, we need to
evaluate the four-point Green's function ($q_1+q_2\to p_1+p_2$):
\begin{eqnarray}
\langle q_1 q_2|p_1 p_2\rangle&\equiv& M(q_1,q_2;p_1,p_2)\nonumber\\
&=& \E^{-\I \frac{\delta}{\delta h_1^{\alpha\beta}}
    D^{\alpha\beta\gamma\delta} \frac{\delta}{\delta
      h_2^{\gamma\delta}}} \GBN [h_{1\mu\nu}]\,
  \GBN[h_{2\mu\nu}]\biggr|_{h_{1\mu\nu}=0=h_{2\mu\nu}} \label{38}\\
&=& -4m_1 m_2 \int d^4x_1\, d^4x_2\, \E^{-\I(q_1-p_1)x_1}
\E^{-\I(q_2-p_2)x_2}\nonumber\\
&&\times\left( \frac{\partial}{\partial
    \tau_1}\frac{\partial}{\partial \tau_2} \E^{\I \frac{1}{2m_1}
  \frac{1}{2m_2} q_2^\alpha q_2^\beta q_1^\gamma q_1^\delta 
  \int\limits_{-\infty}^{\tau_1} d\tau_1'  
  \int\limits_{-\infty}^{\tau_2} d\tau_2'\,
  D_{\alpha\beta\gamma\delta} \left(x_2-x_1+ \frac{q_2}{m_2} \tau_2'
    -\frac{q_1}{m_1} \tau_1'  \right)}
\right)_{\tau_1=0=\tau_2}\!\!\!\!\!\!\!\!\!\!\!\!\!\!\!\!. \label{39}
\end{eqnarray} Since we are mainly interested in the pole structure of
the scattering amplitude, we drop the non-spin-flip factors
$\delta_{\alpha\beta} \delta_{\gamma\delta}$. Expression \re{39} can
best be calculated in the CM system, where it yields ($q=q_1-p_1$,
$q'=q_2 -p_2$): 
\begin{eqnarray}
M(q,q')&=& -4m_1m_2 (2\pi)^4\, \delta(q+q')\int d^4\xi\, \E^{\I q\xi}
\nonumber\\
&&\times \left( \frac{\partial}{\partial
    \tau_1}\frac{\partial}{\partial \tau_2} \E^{\frac{\I}{4m_1m_2}
    16\pi G \bigl( 2(q_1q_2)^2 -m^4\bigr) 
  \int\limits_{-\infty}^{\tau_1} d\tau_1'  
  \int\limits_{-\infty}^{\tau_2} d\tau_2'\,
  D\left( \xi +\frac{q_1}{m_1} \tau_1' -\frac{q_2}{m_2}
    \tau_2'\right)}
\right)_{\tau_1=0=\tau_2}\!\!\!\!\!\!\!\!\!\!\!\!\!\!\!\!. \label{40} 
\end{eqnarray}
In the CM system, $\mathbf{q_2=-q_1}$, we have $s=-(q_1+q_2)^2 =2m^2
-2q_1q_2$, so that
\begin{equation}
2(q_1q_2)^2-m^4 =\frac{1}{2} [(s-2m^2)^2 -2m^4] =:
\gamma(s). \label{41}
\end{equation}
We also introduce four-vectors,
\begin{eqnarray}
u_{1,2}^\mu &=& \frac{1}{\sqrt{2}} (1,0,0,\pm1), \qquad u_1^2=0=u_2^2,
  \qquad u_{1\mu} u_2^\mu =-1, \nonumber\\
\text{so that}&&q_1^\mu q_{2\mu} =-\frac{s}{2}, \label{42}
\end{eqnarray}
following from
\begin{eqnarray}
q_1^\mu&=& \bigl( \sqrt{\mathbf{q_1}^2+m_1^2},0,0,q_1\bigr)
\stackrel{\mathbf{q}_1\to \infty}{\to} (q_1,0,0,q_1)
=\frac{\sqrt{s}}{2} (1,0,0,1), \nonumber\\
q_2^\mu&=& \bigl( \sqrt{\mathbf{q_1}^2+m_2^2},0,0,-q_1\bigr)
\stackrel{\mathbf{q}_1\to \infty}{\to} (q_1,0,0,-q_1)
=\frac{\sqrt{s}}{2} (1,0,0,-1). \nonumber
\end{eqnarray}
A decomposition of $\xi^\mu$ in $\xi^\mu =\xi_{\text{T}}^\mu +\xi_1
u_1^\mu +\xi_2 u_2^\mu$, where the components $\xi_{\text{T}}^\mu$ are
restricted to the $x$-$y$ plane, combined with some more changes of
variables gives
\begin{eqnarray}
M(q,q')&=& -2s(2\pi)^4\, \delta(q+q')\int_{-\infty}^\infty
d^2\boldsymbol{\xi}_{\text{T}}\, \E^{- \I
  \mathbf{q}_{\text{T}}\vec{\xi}_{\text{T}} } \nonumber\\
&&\times \left( \E^{\I \frac{\gamma(s)}{s} 8\pi G
    \int\limits_{-\infty}^\infty d\sigma_1
    \int\limits_{-\infty}^\infty d\sigma_2\, D (\xiT^\mu +u_1^\mu
    \sigma_1 -u_2^\mu \sigma_2)} -1 \right). \label{43}
\end{eqnarray}
Here we introduce a graviton mass $\mu$; then
\begin{displaymath}
D(\xiT +u_1 \sigma_1 -u_2 \sigma_2) =\int \frac{d^4k}{(2\pi)^4} \,
\E^{\I k_\mu \xiT^\mu +\I k_\mu u_1^\mu \sigma_1 -\I k_\mu u_2^\mu
  \sigma_2} \frac{1}{k^2 +\mu^2 -\I \epsilon}. 
\end{displaymath}
Integration over $\int\limits_{-\infty}^\infty d\sigma_1
    \int\limits_{-\infty}^\infty d\sigma_2$ yields two
    $\delta$-functions:
\begin{eqnarray}
\int\limits_{-\infty}^\infty d\sigma_1\, \E^{\I(ku_1)\sigma_1} &=&
2\pi\,\delta(ku_1), \nonumber\\
\int\limits_{-\infty}^\infty d\sigma_2\, \E^{\I(ku_2)\sigma_2} &=&
2\pi\, \delta(ku_2). \nonumber
\end{eqnarray}
But $ku_1=\frac{1}{\sqrt{2}} (k_3-k^0)$, $ku_2=\frac{1}{\sqrt{2}}
(-k_3-k^0)$, so that with the aid of the $\delta$-functions we find:
\begin{displaymath}
k_3=k^0\quad \text{and} \quad k_3=-k^0, \qquad \text{meaning}\quad
k_3=0=k^0.
\end{displaymath}
These facts enable us to write for the exponential in Eq. \re{43}:
\begin{eqnarray}
&&\!\!\!\! \I \frac{\gamma(s)}{s} 8\pi G \int
\frac{d^4k}{(2\pi)^4} \, \frac{\E^{\I k_\mu \xiT^\mu}}{k^2 +\mu^2} \,
(2\pi)^2\, \delta(ku_1)\, \delta(ku_2) \nonumber\\
&&=\I \frac{\gamma(s)}{s} 8\pi G \int
\frac{d^2\mathbf{k}_{\text{T}}}{(2\pi)^2} \, \frac{\E^{\I
    \mathbf{k}_{\text{T}}
    \cdot\vec{\xi}_{\text{T}}^\mu}}{\mathbf{k}_{\text{T}}^2 
  +\mu^2}. \nonumber
\end{eqnarray}
So far we have
\begin{eqnarray}
M(q,q')&=& -2s(2\pi)^4\, \delta(q+q')\int_{-\infty}^\infty
d^2\boldsymbol{\xi}_{\text{T}}\, \E^{- \I
  \mathbf{q}_{\text{T}}\vec{\xi}_{\text{T}} } 
\left( \E^{\I \frac{\gamma(s)}{s} 8\pi G
    \int\frac{d^2k_{\text{T}}}{(2\pi)^2} \, \frac{\E^{\I
    \mathbf{k}_{\text{T}}
    \cdot\vec{\xi}_{\text{T}}^\mu}}{\mathbf{k}_{\text{T}}^2  +\mu^2}}
    -1 \right). \label{44}
\end{eqnarray}
Noticing that
\begin{displaymath}
\frac{K_0(\mu|\mathbf{x}_{\text{T}}|)}{2\pi}
    =\int\frac{d^2\mathbf{k}_{\text{T}}}{(2\pi)^2} \, \frac{\E^{-\I
    \mathbf{k}_{\text{T}} 
    \cdot\vec{\xi}_{\text{T}}^\mu}}{\mathbf{k}_{\text{T}}^2  +\mu^2}, 
\end{displaymath}
and introducing the $T$ matrix via $M=\mathbbm{1} -\I (2\pi)^4\,
\delta(q+q') T$, we obtain
\begin{equation}
\I T=2s \int d^2 \mathbf{x}_{\text{T}}\, \E^{-\I \mathbf{q}_{\text{T}}
  \mathbf{x}_{\text{T}}} \left( \E^{\I \frac{\gamma(s)}{s} 4 G \,
  K_0(\mu|\mathbf{x}_{\text{T}}|) }-1 \right). \label{45}
\end{equation}
The integral over $\mathbf{x}_{\text{T}}$ may be carried out, leading
to 
\begin{equation}
\I T=\frac{2\pi \sqrt{s(s-4m^2)}}{\mu^2} \, \frac{\Gamma\left( 1-\I 2G
    \frac{\gamma(s)}{s} \right)}{\Gamma\left(\I 2G
    \frac{\gamma(s)}{s} \right)} \left( \frac{4\mu^2}{k_{\text{T}}^2}
    \right)^{1-\I 2G \frac{\gamma(s)}{s}}, \label{46}
\end{equation}
or, introducing 
\begin{equation}
\alpha(s)=G \frac{2\gamma(s)}{\sqrt{s(s-4m^2)}}, \label{47}
\end{equation}
\begin{eqnarray}
\I T&=&\frac{2\pi}{\mu^2} \sqrt{s(s-4m^2)} \, \frac{\Gamma\left( 1-\I
    \alpha(s) \right)}{\Gamma\left(\I \alpha(s) \right) }
 \left( \frac{4\mu^2}{-t}
    \right)^{1-\I \alpha(s)},\nonumber\\
&=& \I \frac{16\pi G \gamma(s)}{-t} \frac{\Gamma\left( 1-\I
    \alpha(s) \right)}{\Gamma\left(1+\I \alpha(s) \right) }
 \left( \frac{4\mu^2}{-t}\right)^{-\I \alpha(s)}. \label{48}
\end{eqnarray}
This expression is evidently just the Born amplitude multiplied by 
\begin{displaymath}
\frac{\Gamma\left( 1-\I
    \alpha(s) \right)}{\Gamma\left(1+\I \alpha(s) \right)} 
 \left( \frac{4\mu^2}{-t}\right)^{-\I \alpha(s)}.
\end{displaymath}
Provided that $s>0$ (in fact we are interested in $s\to\infty$), and
$t$ is fixed with $\frac{t}{s}\to 0$, we may set $m=0$ in Eq. \re{47}.
This then brings us to 't Hooft's scattering amplitude \cite{10} in
terms of Lorentz invariant Mandelstam variables:
\begin{equation}
T(s,t)=\frac{8\pi G s^2}{ -t} \frac{\Gamma (1-\I Gs)}{ \Gamma(1+\I
  Gs)} \left( \frac{4\mu^2}{-t} \right)^{-\I Gs}. \label{49}
\end{equation}
The 't Hooft poles occur at $\I Gs=N$, $N=1,2,3,\dots$ These are the
bound state poles in the $\frac{1}{r}$-potential of linearized
gravity. This result had to be expected from similar calculations in
QED. In fact all our results would have come out by just replacing
$eA^\mu$ by $\frac{1}{2} p_\nu h^{\mu\nu}$ in the original action
expression and Green's function equation for spin $1/2$ QED. Hence the
't Hooft poles could have been already discovered by the end of the
sixties, if one had not worried about the non-renormalizability of
Einstein's theory and treated it instead as an effective field theory.

Note that nowhere in this article have we computed closed-loop
processes, in particular, graviton loops. Hence our approximation is
not able to produce any Callan--Symanzik $\beta$-function.
Consequently we cannot generate any quantum correction to Newton's
potential as has been done in more recent articles on the matter
\cite{11} \cite{12}. But this does not come as a surprise since the
eikonal approximation applied to QED is likewise unable to produce
the Uehling potential or Lamb-shift. Finally we should point out that
the 't Hooft poles originate from $|\mathbf{x}_{\text{T}}|$, the
impact parameter, reaching all the way down to zero, and thus from the
short-distance behavior of the Newton interaction. If this is
softened, as, for instance, in the string approach of graviton
scattering, the poles disappear \cite{13}.

\section*{Acknowledgement}
I thank H. Gies for useful discussions and carefully reading the
manuscript.

\end{document}